\documentclass[axioms,article,accept,pdftex,moreauthors]{Definitions/mdpi} 

\firstpage{1} 
\pubvolume{13}
\issuenum{05}
\articlenumber{0323}
\pubyear{2024}
\copyrightyear{2024}
\externaleditor{Academic Editor: Angel Ricardo Plastino}
\datereceived{25 January 2024 } 
\daterevised{3 May 2024 } % Comment out if no revised date
\dateaccepted{9 May 2024 } 
\datepublished{13 May 2024} 
\hreflink{https://doi.org/10.3390/\linebreak axioms13050323} % If needed use \linebreak
\usepackage[utf8]{inputenc} % allow utf-8 input
\usepackage[T1]{fontenc}    % use 8-bit T1 fonts
\usepackage{url}            % simple URL typesetting
\usepackage{booktabs}       % professional-quality tables
\usepackage{amsfonts}       % blackboard math symbols
\usepackage{nicefrac}       % compact symbols for 1/2, etc.
\usepackage{microtype}      % microtypography
\usepackage{xcolor}         % colors

\usepackage{graphicx}
\usepackage{caption}
\usepackage{subcaption}
\usepackage{tikz}  % diagrams using tikz
\usepackage[braket, qm]{qcircuit}
\usepackage{amsmath}
\usepackage{braket}

\usepackage[markup=underlined,final]{changes}
\usepackage{booktabs} % for professional tables
\usepackage{mathtools}
\usepackage{mathrsfs}

\newcommand{\beq}{\begin{equation}}
\newcommand{\eeq}{\end{equation}}
\newcommand{\ga}{\lower.7ex\hbox{$\;\stackrel{\textstyle>}{\sim}\;$}}
\newcommand{\la}{\lower.7ex\hbox{$\;\stackrel{\textstyle<}{\sim}\;$}}

\Title{{Quantum} %MDPI: Dear Authors,(1) Please proofread the article format according to the comments left and pay attention to the highlighted parts.(2) Kindly reply to all comments and do not delete them, since we need to check point by point after receiving your proofed version.(3) Please just modify or confirm or give feedback with track change Function. (4) English editing and layout have been completed, please do not revert the changes..
 Vision Transformers for Quark--Gluon Classification}
\TitleCitation{Quantum Vision Transformers for Quark--Gluon Classification}

\Author{
{Marçal Comajoan Cara} %MDPI: Please carefully check the accuracy of names, emails and affiliations. Authorship can not be changed during this period. Changing authorship is NOT acceptable during this period (including adding new authors/corresponding authors/affiliations or deleting present authors/corresponding authors/affiliations or exchanging author orders)..
 $^{1}$\orcidM{}, 
Gopal Ramesh Dahale $^{2,\dagger}$\orcidG{},
Zhongtian Dong $^{3,\dagger}$\orcidZ{},
Roy T.~Forestano $^{4,\dagger}$\orcidA{},  
\linebreak  Sergei Gleyzer $^{5,\dagger}$\orcidS{},  
Daniel Justice $^{6,\dagger}$\orcidI{},
Kyoungchul Kong $^{3,\dagger}$\orcidK{},  
Tom Magorsch $^{7,\dagger}$\orcidT{},
Konstantin T.~Matchev $^{4,}$*$^{,\dagger}$\orcidB{}, 
Katia Matcheva $^{4,\dagger}$\orcidC{} 
and 
Eyup B.~Unlu $^{4,\dagger}$\orcidD{} }

\AuthorNames{Marçal Comajoan Cara, Gopal Ramesh Dahale, Zhongtian Dong, Roy Forestano, Sergei Gleyzer, Daniel Justice, Kyoungchul Kong, Tom Magorsch, Konstantin Matchev, Katia Matcheva and Eyup Unlu}

\AuthorCitation{Comajoan Cara, M.; Dahale, G.R.; Dong, Z.; Forestano, R.T.; Gleyzer, S.; Justice, D.; Kong, K.; Magorsch, T.; Matchev, K.T.; Matcheva, K.; et al.}
\address{%
$^{1}$ \quad  Department of Signal Theory and Communications, Polytechnic {University} %MDPI: IMPORTANT: For universities, the department/school/faculty/campus is required. Please try to provide this information.
 of Catalonia,
\linebreak08034 Barcelona, Spain; {marcal.comajoan@estudiantat.upc.edu} %MDPI: We added the email according the system, please check and confirm. The same as following
\\
$^{2}$ \quad Indian Institute of Technology Bhilai,
  Bhilai 491001, {Chhattisgarh}, India; {gopald@iitbhilai.ac.in}\\
$^{3}$ \quad Department of Physics and Astronomy,
  University of Kansas,
  Lawrence, KS 66045, USA; \linebreak  {cdong@ku.edu (Z.D.); kckong@ku.edu (K.K.)} \\  
$^{4}$ \quad Institute for Fundamental Theory, Physics Department, University of Florida, Gainesville, FL 32611, USA; {roy.forestano@ufl.edu (R.T.F.); matcheva@ufl.edu (K.M.); eyup.unlu@ufl.edu (E.B.U.)}\\
$^{5}$ \quad Department of Physics and Astronomy,
  University of Alabama,
  Tuscaloosa, AL 35401, USA; {sgleyzer@ua.edu}\\
$^{6}$ \quad Software Engineering Institute, Carnegie Mellon University,  Pittsburgh, PA 15213, USA; {dljustice@sei.cmu.edu} \\
$^{7}$ \quad Physik-Department, Technische {Universit\"at} %MDPI: We revised to full name, please confirm
 München, 85748 Garching, Germany; {tom.magorsch@tum.de}}

\corres{Correspondence: matchev@ufl.edu}
\firstnote{These authors contributed equally to this work.}
\abstract{We \replaced{introduce}{present} a hybrid quantum-classical vision transformer architecture, \replaced{notable for its integration of variational quantum circuits within both the attention mechanism and the multi-layer perceptrons.  The research addresses the critical challenge of computational efficiency and resource constraints in analyzing data from the upcoming High Luminosity Large Hadron Collider, presenting the architecture as a potential solution. In particular, we evaluate our method by applying the model to multi-detector jet images from CMS Open Data.}{which employs variational quantum circuits to compute the required matrices in the attention calculations, as well as in the multi-layer perceptrons.  We apply the model to multi-detector jet images from the simulated CMS Open Data.} The goal is to distinguish quark-initiated from gluon-initiated jets. We successfully train the quantum model and evaluate it via numerical simulations. Using this approach, we achieve classification performance almost on par with the one obtained with the completely classical architecture, considering a similar number of parameters.}



\begin{document}

\section{Introduction}

\replaced{The imminent operation of the High Luminosity Large Hadron Collider (HL-LHC)~\cite{hllhc} by the end of this decade signals an era of unprecedented data generation, necessitating vast computing resources and advanced computational strategies to effectively manage and analyze the resulting datasets~\cite{HSFPhysicsEventGeneratorWG:2020gxw}.}{The High Luminosity Large Hadron Collider (HL-LHC) is planned to start its operation at the end of this decade. The program will produce enormous quantities of data, which in turn will require vast computing resources~\cite{hllhc}.} 
A promising approach to deal with this huge amount of data could be the application of quantum machine learning (QML), which could reduce the time complexity of classical algorithms by running on quantum computers and obtain better accuracies thanks to the access to the exponentially large Hilbert space ~\cite{qml1,qml2,hilbert,qml3,qml4,qml5,qml6,axioms13030188_Cosmos,axioms13030160_Roy}.

% \deleted{The main idea behind QML is to use models that are partially or fully executed on a quantum computer by replacing some subroutines of the models with quantum circuits \added{in order to exploit the unique properties of quantum mechanics to enhance the capabilities of classical machine learning algorithms}. Some \added{notable} examples are quantum support vector machines~\cite{qsvm}, quantum nearest-neighbor algorithms~\cite{qknn}, quantum nearest centroid classifiers~\cite{qnc}, or quantum artificial neural networks~\cite{qml3,axioms13030188_Cosmos}, \added{including quantum graph neural networks~\cite{axioms13030160_Roy}}. In the last case, typically some layers are executed on a quantum circuit which has rotation angles that are free parameters of the whole model. These parameters are optimized together with the parameters of the classical layers. Such parametrized quantum circuits are also called variational quantum circuits (VQCs).}

\added{The innovative core of our research lies in the development of a novel quantum-classical hybrid vision transformer architecture that integrates variational quantum circuits into the attention mechanisms and multi-layer perceptions of the classical vision transformer (ViT) architecture~\cite{vit}.} \replaced{More specifically,}{In this work,} we adapt the classical ViT architecture to the quantum realm by replacing the classical linear projection layers used in the multi-head attention subroutines by \replaced{variational quantum circuits (VQCs)}{VQCs}, as well as by using VQCs in the multi-layer perceptrons too. This approach is based on previous work~\cite{disipio}, which proposed the same idea for the original transformer architecture for text~\cite{transformer}. Other works have explored other possible quantum adaptations of the original transformer~\cite{disipio,li}, as well as adaptations of the vision transformer~\cite{cherrat,axioms13030187_Eyup} and the graph transformer~\cite{qgt}. \added{Our work differs from~\cite{cherrat} in the architecture that we propose, which explores the use of other quantum ansatzes. The model in~\cite{axioms13030187_Eyup} was developed in parallel with this study and differs in several respects---the use of classical multi-layer perceptrons (MLPs) instead of quantum MLPs  and the use of different ansatzes for the key, value, and query operations.}

We train and evaluate \replaced{our proposed}{the} quantum vision transformer (QViT) on multi-detector jet images from data from the CMS Open Data Portal~\cite{cmsopendata}.  The goal is to discriminate between quark-initiated (quark) and gluon-initiated (gluon) jets. \replaced{This}{Such} task has broad applicability to searches and measurements at the Large Hadron Collider (LHC) ~\cite{qgatlas}. Consequently, ways of solving this  this task have already been extensively examined with classical machine learning techniques~\cite{qgatlas,qgcms,qg1,qg2,qggleyzer}.

\added{The motivation behind the application of QML to this particular task stems from the inherent limitations of current classical deep learning models, which, despite their efficacy, are increasingly constrained by escalating computational demands and resource requirements inherent in processing and analyzing large datasets, such as those anticipated from the HL-LHC. Our research endeavors to address these challenges by leveraging the unique capabilities of quantum computing to enhance the efficiency and performance of machine learning models in the context of high-energy physics.}

\section{\added{Background}}

\subsection{\added{(Classical) Deep Learning, the Transformer, and the Vision Transformer}}

\added{The field of artificial intelligence aims to replicate in computers the remarkable capabilities of the human brain, such as identifying objects in images, writing text, transcribing and recognizing speech, offering personalized recommendations, and much more. The application of machine learning systems is becoming ubiquitous in many domains of science, technology, business, and government, gradually replacing the use of traditional hand-crafted algorithms. This shift has not only enhanced the efficacy of existing technologies but has also paved the way for an array of novel capabilities that would have been inconceivable otherwise.}

\added{Deep learning is a subfield of artificial intelligence that deals with neural networks, a type of computational model that has emerged as an exceptionally powerful and versatile approach to learning from data. The most straightforward realization of a neural network is in a ``feedforward'' configuration, also known as a multi-layer perceptron (MLP), which can be mathematically described as a composition of elementwise non-linearities with affine transformations of the data}~\cite{lbh,bishop,sh}.

\added{In this context, an affine transformation refers to a linear transformation followed by a translation. Given an input vector $x \in \mathbb{R}^{D_1}$, a weight matrix $W \in \mathbb{R}^{D_2 \times D_1}$, and a bias vector $b \in \mathbb{R}^{D_2}$, the affine transformation is defined as}
\begin{equation}
\added{a(x) = Wx + b} \, ,
\end{equation}
\added{where $a(x) \in \mathbb{R}^{D_2}$ is the output of the affine transformation.}

\added{The elementwise non-linearity, also known as an activation function, is then applied to each component of the output vector $a$:}
\begin{equation}
\added{f(x) = \sigma(a(x))} \, ,
\end{equation}
\added{where $\sigma$ denotes the activation function. Traditional choices for the activation function include the sigmoid function and the hyperbolic tangent (tanh) function, but these have largely fallen out of favor in modern deep learning architectures. The rectified linear unit (ReLU)~\cite{relu1}, defined as}
\begin{equation}
\added{\mathrm{ReLU}(x) = \max(0, x),}
\end{equation}
\added{has gained popularity due to its simplicity and effectiveness~\cite{relu2}. More recently, variations of the ReLU have been proposed to further improve the performance and stability of deep learning models, such as the Gaussian Error Linear Unit (GELU)~\cite{gelu}, which is defined as}
\begin{equation}
\added{\mathrm{GELU}(x) = x \Phi(x),}
\end{equation} 
\added{where $\Phi(x)$ is the cumulative distribution function of the standard normal distribution. Another important activation function in deep learning is the softmax function, which is commonly used in the output layer of a neural network for multi-class classification tasks. The softmax function takes a vector of real numbers and transforms it into a probability distribution over the classes. Given an input vector $z \in \mathbb{R}^{K}$, the softmax function is defined~as }
\begin{equation}
\added{\text{softmax}(z_i) = \frac{e^{z_i}}{\sum_{j=1}^{K} e^{z_j}}. }
\end{equation} 
\added{The output of the softmax function represents the predicted probabilities for each class, with the highest probability indicating the most likely class.}

\added{Deep learning networks are constructed by stacking multiple layers of these transformations:}
\begin{equation}
\added{\hat{y} = f_L \circ f_{L-1} \circ \cdots \circ f_1(x)} \, ,
\end{equation}
\added{where} 
\begin{equation}
\added{f_i(x) = \sigma_i(W_ix+b_i).}
\end{equation} 
\added{This stacking allows the network to learn increasingly complex representations of the input data. The output of one layer serves as the input to the subsequent layer, forming a hierarchical structure. The final layer of the network produces the desired output, which can be a classification label, a regression value, or any other task-specific output.}

\added{The learning process in deep learning involves adjusting the weights and biases of the network to minimize a loss function, which quantifies the discrepancy between the predicted outputs and the expected ones. For classification tasks, a commonly used loss function is the cross-entropy loss, which measures the dissimilarity between the predicted class probabilities and the true class labels. The cross-entropy loss is defined as}
\begin{equation}
\added{L = -\sum_{n=1}^{N} \sum_{k=1}^{K} y_{nk} \log(\hat{y}_{nk})} \, ,
\end{equation}
\added{where $N$ is the number of samples, $K$ is the number of classes, $y_{nk}$ is the true label (0 or 1) for sample $n$ and class $k$, and $\hat{y}_{nk}$ is the predicted probability for sample $n$ and class $k$.}

\added{The optimization of the loss function is typically performed using stochastic gradient descent (SGD) or its variants. SGD updates the model parameters using a randomly selected subset of the training data, called a mini-batch, at each iteration. The update rule for SGD is given by}
\begin{equation}
\added{\theta_{t+1} = \theta_t - \eta \nabla_{\theta} L_B(\theta_t)} \, ,
\end{equation}
\added{where $\theta_t$ represents the model parameters at iteration $t$, $\eta$ is the learning rate, and $\nabla_{\theta} L_B(\theta_t)$ is the gradient of the loss function with respect to the parameters, estimated using the mini-batch $B$.}

\added{The backpropagation algorithm is typically used to efficiently compute the gradients of the loss function with respect to the model parameters in a neural network. It relies on the chain rule of calculus to propagate the gradients from the output layer to the input layer, enabling the computation of the gradients for each layer in the network.}

\added{Apart from the MLP, more advanced neural network architectures have been devised. Among these, the Transformer architecture~\cite{transformer} stands out as a seminal breakthrough in the field of deep learning. The main building block of the Transformer is a layer that takes as input a matrix $X\in\mathbb{R}^{N\times D}$ and outputs a transformed matrix $X'\in\mathbb{R}^{N\times D}$ of the same dimensionality. Each of these layers has two sub-layers: first, a multi-head self-attention mechanism, the core architectural component of the Transformer, and second, a simple MLP. Moreover, to improve training efficiency,  layer normalization~\cite{layernorm} and residual connections~\cite{resnet} around each sub-layer are employed. Thus, the resulting transformation~is}
\begin{gather}
    \added{Z = X+\mathrm{LayerNorm}(\mathrm{MHA}(X,X,X)),}\\
    \added{X' = Z+\mathrm{LayerNorm}(\mathrm{MLP}(Z))} \, .
\end{gather}

\added{The attention mechanism is a key component of the Transformer architecture. It allows the model to focus on specific parts of the input sequence when generating each output element. Given a query matrix $Q\in\mathbb{R}^{N\times D_k}$, a key matrix $K\in\mathbb{R}^{M\times D_k}$, and a value matrix $V\in\mathbb{R}^{M\times D_v}$, the attention function is defined in~\cite{transformer} as:}
\begin{equation}
    \mathrm{Attention}(Q, K, V) = \mathrm{softmax}\left(\frac{QK^T}{\sqrt{D_k}}\right)V \, ,
\end{equation}
\added{where $D_k$ is the dimension of the keys, used as a scaling factor to prevent the dot products from growing too large.}

\added{Self-attention is a special case of attention where the query, key, and value matrices are all derived from the same input matrix $X$. In the Transformer, self-attention allows each position in the input sequence to attend to all positions in the previous layer.}

\added{Multi-head attention is an extension of the attention mechanism that allows the model to jointly attend to information from different representation subspaces at different positions. Instead of performing a single attention function, multi-head attention linearly projects the queries, keys, and values $h$ times with different learned linear projections, performs the attention function in parallel, concatenates the results, and projects the concatenated output using another learned linear projection. Mathematically, multi-head attention is defined as}
\begin{equation}
\begin{split}
    \mathrm{MHA}(Q, K, V) &= \mathrm{Concat(head_1, ..., head_h)}W^O,\\
    \text{where } \mathrm{head_i} &= \mathrm{Attention}(QW_i^Q, KW_i^K, VW_i^V)
\end{split}
\label{eq:mha}
\end{equation}
where $W_i^Q\in\mathbb{R}^{D\times D_k}$, $W_i^K\in\mathbb{R}^{D\times D_k}$, $W_i^V\in\mathbb{R}^{D\times D_v}$, and $W^O\in\mathbb{R}^{hD_v\times D}$ are learnable parameter matrices.

\added{The Transformer architecture, originally designed for natural language processing, has also been adapted to other domains. For instance, its adaptation for computer vision has given rise to the Vision Transformer (ViT)~\cite{vit}. In ViTs, an image is split into a sequence of patches, which are then linearly embedded and treated as input tokens for a stack of Transformer layers, collectively referred to as the Transformer encoder. The ViT has achieved state-of-the-art performance on various image classification benchmarks, demonstrating the versatility and effectiveness of the Transformer architecture across different domains~\cite{bettervit}.}

\subsection{\added{Quantum Computing and Quantum Machine Learning}}

\added{In quantum computing, the fundamental unit of information is the qubit, which, unlike its classical counterpart, the bit, can exist in a state of superposition to represent non-binary states. The quantum state of $n$ qubits can be represented with a unit vector $\ket{\psi}$ in the Hilbert space $\mathbb{C}^{2^n}$ (in bra-ket notation, the ket $\ket{~}$ denotes a column vector and the bra $\bra{~}$ a row vector).}

\added{A quantum circuit is a series of quantum logic operations (or gates) applied to qubits to change their state. This can be represented mathematically by matrix multiplication, $U\ket{\psi}$, where $U$ is a $2^n\times2^n$ unitary matrix. Typically, a quantum circuit ends with a measurement of all the qubits, which provides important information about the final state of the circuit.}

\added{In this paper, we make use of the $R_X$ gate, which performs a single-qubit rotation about the $X$ axis, and the CNOT gate, which operates over two qubits, by flipping the second one (the target qubit) if and only if the first one (the control qubit) is $\ket{1}$. They can be represented with the following matrices:}
\begin{eqnarray}
R_X(\theta) &=& 
\begin{bmatrix}
    \cos{(\theta/2)}&-i\sin{(\theta/2)}\\ 
    -i\sin{(\theta/2)}&\cos{(\theta/2)}
\end{bmatrix},\\[2mm]
CNOT &=& 
\begin{bmatrix}
    1&0&0&0\\
    0&1&0&0\\
    0&0&0&1\\
    0&0&1&0\\
\end{bmatrix} \, .
\end{eqnarray}

The main idea behind quantum machine learning (QML) is to use models that are partially or fully executed on a quantum computer by replacing some subroutines of the models with quantum circuits \added{in order to exploit the unique properties of quantum mechanics to enhance the capabilities of classical machine learning algorithms}. Some \added{notable} examples are quantum support vector machines~\cite{qsvm}, quantum nearest-neighbor algorithms~\cite{qknn}, quantum nearest centroid classifiers~\cite{qnc}, and quantum artificial neural networks~\cite{qml3,axioms13030188_Cosmos}, \added{including quantum graph neural networks~\cite{axioms13030160_Roy}}. In the last case,  some layers are typically executed on a quantum circuit that has rotation angles that are free parameters of the whole model. These parameters are optimized together with the parameters of the classical layers. Such parametrized quantum circuits are also called variational quantum circuits (VQCs).

\subsection{\added{High-Energy Physics and Jets}}

\added{
High-energy physics research aims to understand how our universe works at its most fundamental level. We do this by discovering the most elementary constituents of matter and energy, exploring the basic nature of space and time itself  and probing the interactions between them. These fundamental ideas are at the heart of physics and hence all of the physical sciences. Among many other experiments, the LHC provides ubiquitous opportunities for precision measurement of particle properties in the standard model of elementary particle physics, as well as for searching for new physics beyond the standard model. 
It is not only the largest human-made experiment on Earth but also the most prolific producer of scientific data. The HL-LHC will produce 100-fold to about 1 exabyte per year, bringing quantitatively and qualitatively new challenges due to its event size, data volume, and complexity, therefore straining the available computational resources~\cite{Franceschini:2022vck}. }

\added{In collider experiments, jets arise as a result of the hadronization of the fundamental elementary particles, which carry color charge, namely, the quarks and the gluons. The color confinement phenomenon in quantum chromodynamics implies that quarks and gluons cannot exist in free form  but must be converted into a collection of colorless objects (called hadrons)~\cite{Ellis:1996mzs}. In high-energy particle collisions like those taking place at the LHC, the initial quarks and gluons are produced with significant boosts (i.e., with large momenta), and therefore, the resulting collections of hadrons appear as narrow collimated bunches, which are generically called jets. There are standard and well-tested jet reconstruction algorithms that identify candidate jets among the myriad of particles observed in the detector~\cite{Salam:2010nqg}. However, the question of the precise origin of a given jet---whether it came from a quark (and which type of quark) or a gluon---is highly non-trivial and to this day continues to be the subject of active investigations in the literature}~\cite{Larkoski:2017jix,Kogler:2018hem,Marzani:2019hun}\added{.}

\added{In this paper, we shall focus on the classification task of distinguishing between a jet arising from a light quark and a jet arising from a gluon progenitor particle (see}~\cite{Feickert:2021ajf,Larkoski:2017jix,Kogler:2018hem,Marzani:2019hun,Guest:2018yhq,Albertsson:2018maf,Radovic:2018dip,Carleo:2019ptp,Bourilkov:2019yoi,Schwartz:2021ftp,Karagiorgi:2021ngt,Boehnlein:2021eym,Shanahan:2022ifi}
\added{and references therein for various classical machine learning methods).}

\section{Method}

\subsection{Data}

We use the dataset described in~\citet{qggleyzer}, \added{which was} derived from simulated data \added{for QCD dijet production} available on the CERN CMS Open Data Portal~\cite{cmsopendata}. \added{Events were generated and hadronized with the {\sc PYTHIA6} Monte Carlo event generator using the Z2$^\ast$ tune, which accounts for the difference in the hadronization patterns of quarks and gluons.} The dataset consists of 933,206 3-channel $125 \times 125$ images, with half representing quarks and the other half gluons. Each of the three channels in the images corresponds to a specific component of the Compact Muon Solenoid (CMS) detector~\cite{CMS:2008xjf}: the inner tracking system (Tracks), which identifies charged particle tracks~\cite{CMS:2014pgm}; the electromagnetic calorimeter (ECAL), which captures energy deposits from electromagnetic particles~\cite{CMS:2013lxn}; and the hadronic calorimeter (HCAL), which detects energy deposits from hadrons~\cite{CMSHCAL:2007zcq,CMSHCAL:2008fum}.

\begin{figure}[H]

\includegraphics[width=.9\textwidth]{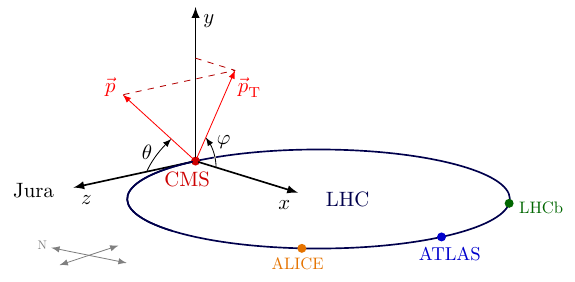}

\caption{{The} %MDPI: We moved all figures after the first mention. Please check and confirm.
 CMS coordinates the system against the backdrop of the LHC, with the location of the four main experiments (CMS, ALICE, ATLAS, and LHCb). The $z$ axis points to the Jura mountains, while the $y$-axis points toward the sky. In spherical coordinates, the components of a particle momentum $\vec{p}$ are its magnitude $|\vec{p}|$, the polar angle $\theta$ (measured from the $z$-axis), and the azimuthal angle $\varphi$ (measured from the $x$-axis). The transverse momentum $\vec{p}_T$ is the projection of $\vec{p}$ on the transverse ($xy$) plane. This figure was generated with TikZ code adapted from Ref.~\cite{CMS_Coordinate_System}.}
\label{fig:CMS}
\end{figure}

\added{In the CMS experiment, the components of the measured momenta of individual particles are represented in a coordinate system oriented as shown in Figure~\ref{fig:CMS}~\cite{CMS:2008xjf}. The origin of the coordinate system is centered at the nominal collision point inside the experiment, the $y$-axis points vertically up, and the x-axis points radially inward toward the center of the LHC. In order to form a right-handed coordinate system, the $z$-axis then points along the beam direction toward the Jura mountains from LHC Point 5 (the location of the CMS experiment). The azimuthal angle $\varphi$ is measured from the $x$-axis in the $(x,y)$ plane, while the polar angle $\theta$ is measured from the $z$-axis. In particle physics, one often trades the polar angle $\theta$ for related kinematic variables like the rapidity $y$ or the closely related pseudorapidity $\eta$, which are defined as~\cite{Franceschini:2022vck} }
\begin{equation}
\added{y \equiv \frac{1}{2} \ln \frac{E + p_z} {E - p_z}}
\end{equation}
\added{and}
\begin{equation}
\added{\eta \equiv - \ln \left[\tan\left(\frac{\theta}{2}\right)\right]. }
\end{equation}
\added{Furthermore, the magnitude of the momentum $\vec{p}_T$ transverse to the beam direction is computed from the respective $p_x$ and $p_y$ components as}
\begin{equation}
\added{ p_T \equiv \sqrt{p_x^2+p_y^2}. }
\end{equation}

\vspace{-12pt}

\begin{figure}[H]
     
     \includegraphics[width=\textwidth]{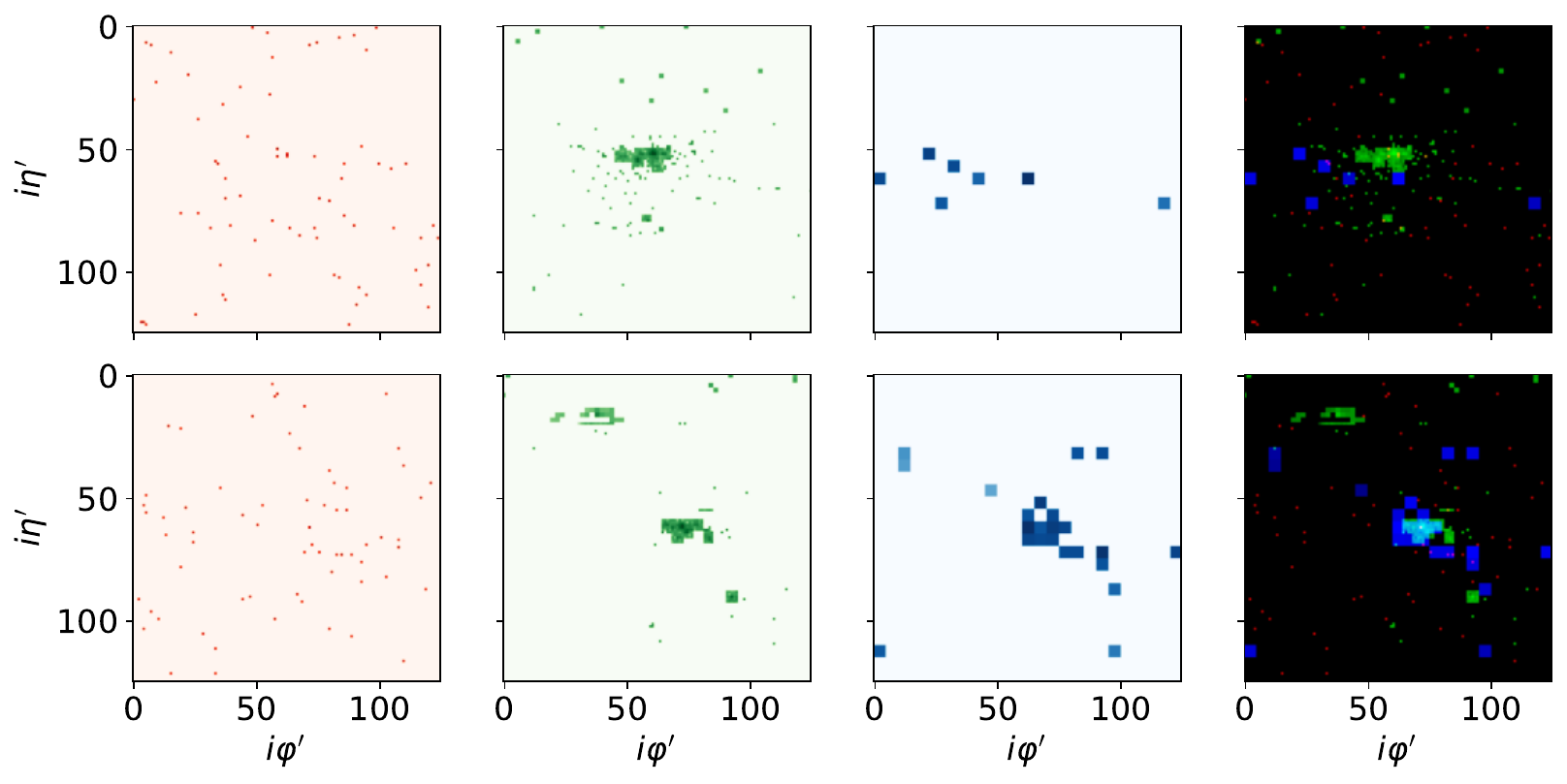}
     \caption{{Representative} %MDPI: Please confirm if different colors need add further explanation.
 images of jets for both quarks (\textbf{\boldmath{top}}) and gluons (\textbf{\boldmath{bottom}}). The columns show the distinct sub-detectors: Tracks, ECAL, HCAL, and a composite image combining all three. All images are in log scale. Note that the ECAL and HCAL were upscaled to match the Tracks resolution.}
     \label{fig:one}
\end{figure}

\begin{figure}[H]
     
     \includegraphics[width=\textwidth]{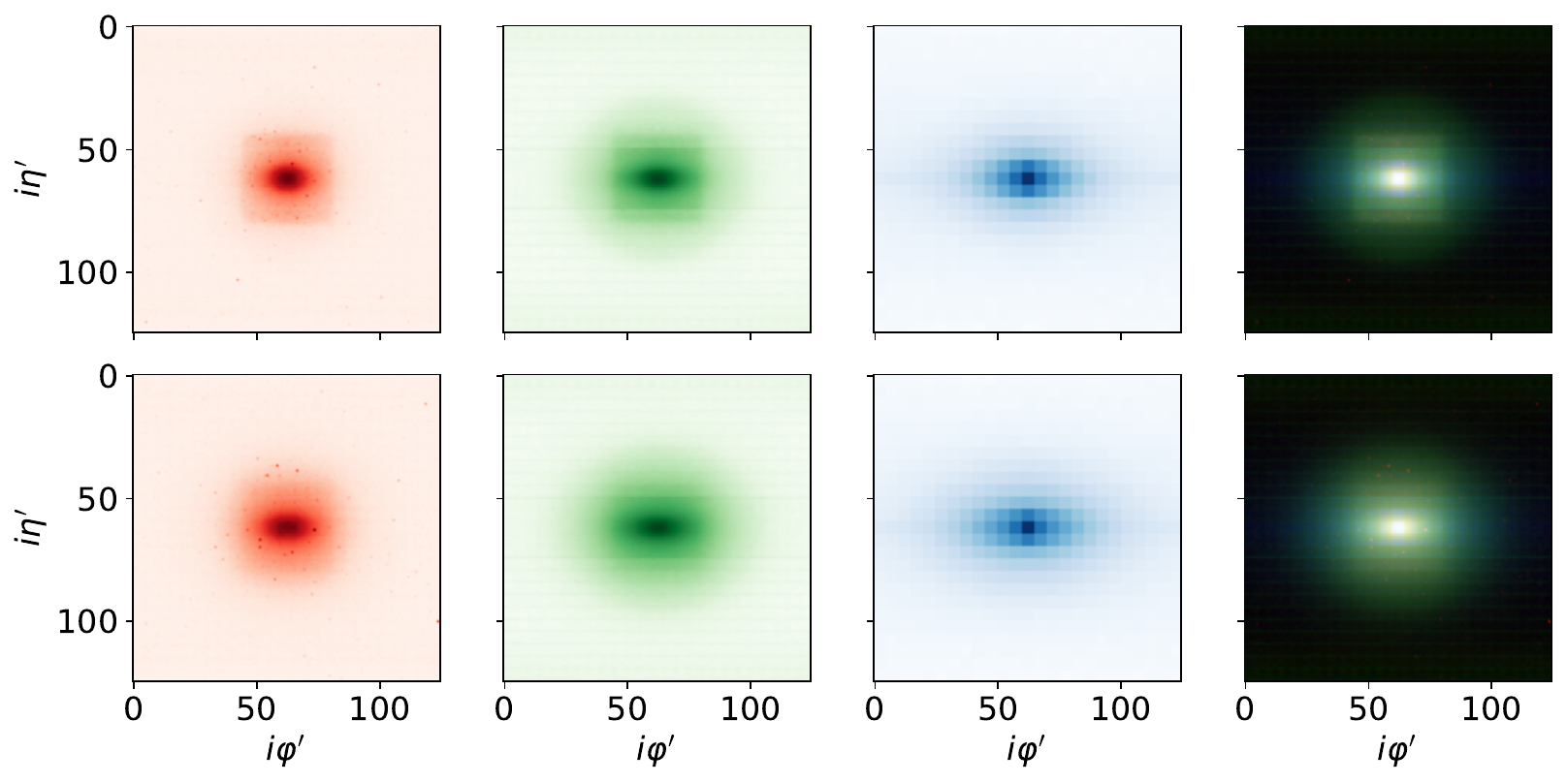}
     \caption{{Average} %MDPI: Please confirm if different colors need add further explanation.
 images of quarks (\textbf{\boldmath{top}}) and gluons (\textbf{\boldmath{bottom}}) across the entire dataset.  The columns show the distinct sub-detectors: Tracks, ECAL, HCAL, and a composite image combining all three. All images are in log scale. Note the more dispersed nature of the gluon jets across channels.}
     \label{fig:avg}
\end{figure}

For a more \replaced{intuitive}{comprehensive} understanding of the jet images \replaced{in our dataset, we show several visualizations in}{see} Figures~\ref{fig:one} and \ref{fig:avg}. \added{Figure~\ref{fig:one} shows the various subdetector images for a single jet: a representative quark jet in the upper row and a representative gluon jet in the bottom row. Then, Figure~\ref{fig:avg} shows the corresponding subdetector images averaged over the full dataset. The ECAL images have $125\times 125$ resolution in the plane of the azimuthal angle $\varphi'$ and the pseudorapidity $\eta'$, while the HCAL resolution is only $25\times 25$ in the $(\varphi', \eta')$ plane.}
\vspace{-3pt}
\vspace{-9pt}

\subsection{Model}

\begin{figure}[H]
\centerline{
\input{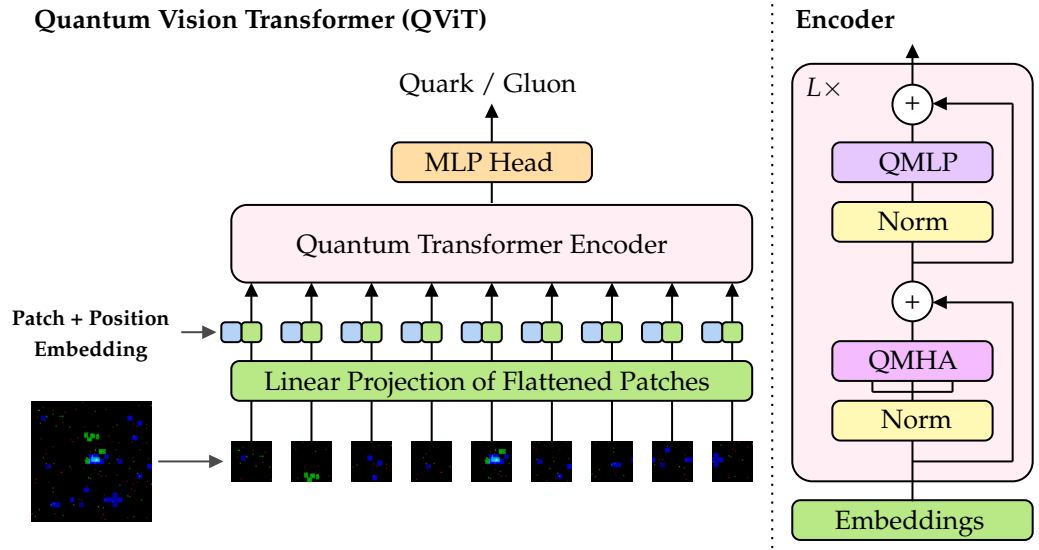}
}
\caption{{Model} %MDPI: Please confirm if different colors need add further explanation.
 overview. QMHA stands for quantum multi-head attention and QMLP for quantum multi-layer perceptron. \replaced{The drawing style of the illustration was}{The illustration was} inspired by~\citet{vit}, the major difference being that here we use a quantum transformer encoder as depicted in the right side of the figure.}
\label{fig:model}
\end{figure}

As in the original classical ViT~\cite{vit}, the image is split into patches that are linearly embedded together with position embeddings. Nonetheless, the change we introduce is that these patches are instead fed to the Quantum Transformer Encoder, which employs VQCs in the multi-head attention (MHA) and multi-layer perceptron (MLP) components. An overview of the model is shown in Figure~\ref{fig:model}.

More concretely, the output of the classical multi-head attention layer is computed \replaced{by using VQCs to compute all four linear projections in the MHA computations (Equation~(\ref{eq:mha})) instead of classical feedforward layers.}{as: [...] Here, instead, we use VQCs to act as the classical fully connected layers.} Similarly, in the MLP component of the encoder, we also employ VQCs to replace classical fully connected layers. \added{Nonetheless, note that the activation functions in the MLP, which are GELU~\cite{gelu}, are executed classically}.

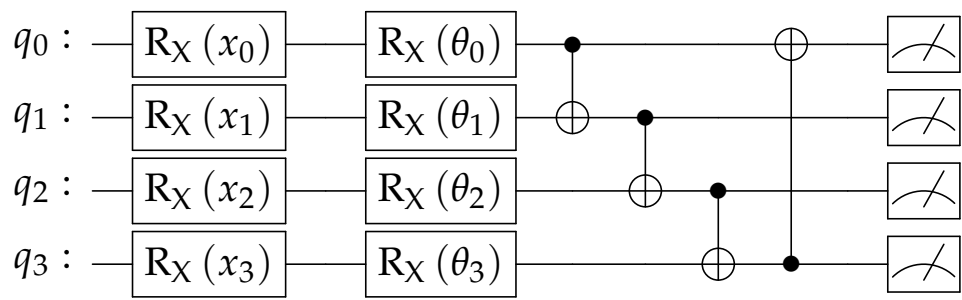
\begin{figure}[H]
    \hspace{-20pt}
    \scalebox{1.5}{
    \Qcircuit @C=1.0em @R=0.2em @!R { \\
    	 	\nghost{{q}_{0} :  } & \lstick{{q}_{0} :  } & \gate{\mathrm{R_X}\,(x_0)} & \qw & \gate{\mathrm{R_X}\,(\theta_0)} & \ctrl{1} & \qw & \qw & \targ & \qw & \meter\\
    	 	\nghost{{q}_{1} :  } & \lstick{{q}_{1} :  } & \gate{\mathrm{R_X}\,(x_1)} & \qw & \gate{\mathrm{R_X}\,(\theta_1)} & \targ & \ctrl{1} & \qw & \qw & \qw & \meter\\
    	 	\nghost{{q}_{2} :  } & \lstick{{q}_{2} :  } & \gate{\mathrm{R_X}\,(x_2)} & \qw & \gate{\mathrm{R_X}\,(\theta_2)} & \qw & \targ & \ctrl{1} & \qw & \qw &\meter\\
    	 	\nghost{{q}_{3} :  } & \lstick{{q}_{3} :  } & \gate{\mathrm{R_X}\,(x_3)} & \qw & \gate{\mathrm{R_X}\,(\theta_3)} & \qw & \qw & \targ & \ctrl{-3} & \qw & \meter\\
    \\ }}
    \vspace{-8pt}
    \caption{Variational quantum circuits used in the proposed QViT.}
    \label{fig:vqc}
\end{figure}

In particular, the VQC configuration we use is the one shown in Figure \ref{fig:vqc}. First, each feature of the vector $x=(x_0, ..., x_{n-1})$ is embedded into the qubits by encoding them into their rotation angles. Next, a layer of one-parameter single-qubit rotations acts on each wire. These parameters, $\theta = (\theta_0, ..., \theta_{n-1})$, are learned together with the rest of the parameters of the model. Then, a ring of CNOT gates follows to entangle the qubit states. Thus, the obtained behavior is similar to a matrix multiplication. Finally, each qubit is measured, and the output is fed to the next corresponding component of the encoder.
\vspace{-12pt}

We train both the proposed QViT and a classical ViT with the same hyperparameters to have a meaningful baseline for comparison. We use a patch size of ten, a hidden size of eight, and four transformer blocks with four attention heads each and a  hidden MLP size of four.

\added{As suggested by recent work on benchmarking quantum utility~\cite{Herrmann_2023}, we choose the classical and the quantum architectures to have a similar number of trainable parameters. Note that since the input and output states of a VQC have the same dimension, the number of qubits has to coincide with the size of the corresponding layers in the neural network. This results in the use of four circuits made up of four qubits for the QMHA layer of each transformer block, and, likewise, four circuits made up of four qubits for the QMLs. In total, the classical ViT has 5178 parameters, while the QViT has 4170 parameters. The smaller number in the QViT is due to the fact that the proposed VQC has only $n$ free parameters, while a classical fully connected layer with bias has $n^2+n$ parameters.}

The dimensions \added{used}  are small so that the circuits do not require many qubits. Consequently, the simulation time is not very long, and the model can be executed in already existing quantum hardware.

We use a batch size of 256 and train for 25 epochs with the AdamW optimizer~\cite{adamw} with gradient clipping at norm 1, and a learning rate scheduler that first performs a linear warmup for 5000 steps from 0 to $10^{-3}$, followed by cosine decay~\cite{cosine}. We execute a random hyperparameter search to find good parameters in the classical baseline and apply them to the QViT.

\replaced{We use the same training--validation--test split as in Andrews et al.~\cite{qggleyzer}. In particular, of the whole dataset, 714,510 images are allocated for training, 79,390 for validation, and 139,306   for the final test set. To assess the classifier's performance, we employ the Receiver Operating Characteristic (ROC) curve. In the context of high-energy physics, this curve can be interpreted in terms of signal efficiency (true positive rate) versus background rejection (true negative rate).}{Of the whole dataset, 714,510 images are allocated for training, 79,390 for validation, and 139,306 are reserved for the final test set.}
The area under the ROC curve (AUC) is computed for each epoch of each model configuration. After all the epochs, we select the parameters from the epoch that achieves the highest validation AUC  and reevaluate them on the separate hold-out test set to obtain the final test AUC.

We use JAX~\cite{jax} and Flax~\cite{flax} to implement the classical parts of the model and the classical baseline, as well as to train both models. We use TensorCircuit~\cite{tensorcircuit} to implement, train, and execute the VQCs by numerical simulation on a classical computer. By using TensorCircuit, we are able to train the quantum model for several epochs in a \replaced{relatively}{relative} short amount of time (\replaced{around 39}{30} minutes per epoch). This is an improvement over previous works, such as~\citet{disipio}, which required about 100 h to train \replaced{a similar hybrid transformer model for just one epoch}{for a single epoch a similar hybrid transformer model}, even though we have many more samples.

\section{Results}

\begin{figure}[H]
     
     \includegraphics[width=0.63\textwidth]{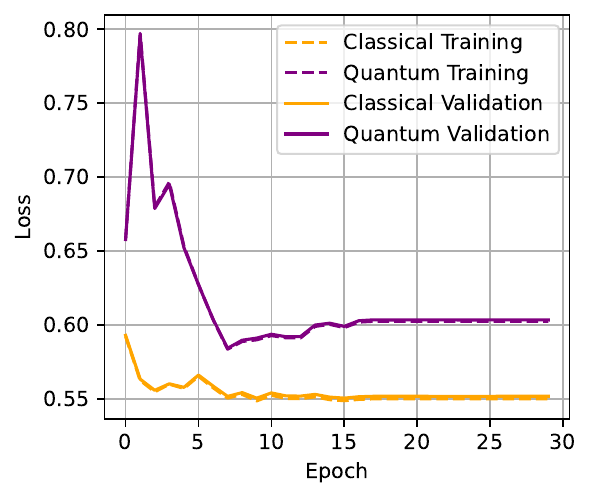}
     \caption{\added{Binary cross-entropy loss evolution during training, computed at the end of each epoch on the training (dashed lines) and validation (solid lines) sets for both the baseline classical ViT (orange lines) and our hybrid QViT (purple lines).}}
     \label{fig:loss}
\end{figure}

\begin{figure}[H]
     
     \includegraphics[width=0.67\textwidth]{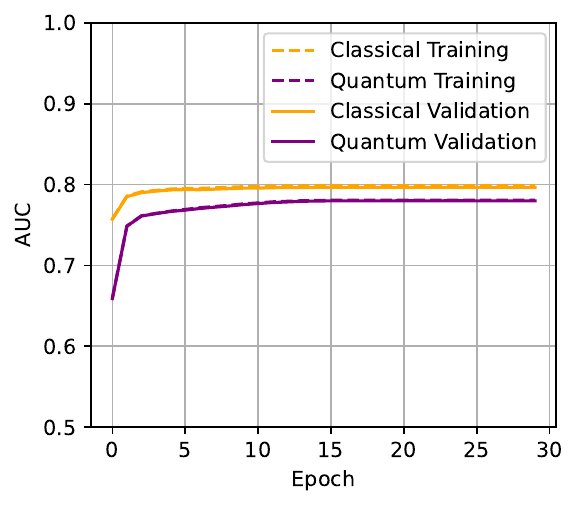}
     \caption{\added{AUC score evolution during training computed at the end of each epoch on the training (dashed lines) and validation (solid lines) sets for both the baseline classical ViT (orange lines) and our hybrid QViT (purple lines).}}
     \label{fig:auc}
\end{figure}

\added{The evolution of the loss and AUC score during training, computed at the end of each epoch, is shown in Figures \ref{fig:loss} and \ref{fig:auc}, respectively. We do not observe signs of overfitting in any case, as the training and validation curves are almost the same.}

\vspace{-9pt}

\added{The epoch that obtains the highest validation AUC is the 16th in the case of the classical ViT and   25th in the case of our hybrid QViT. Although the ViT converges faster, we observe that the QViT converges quite fast too, but keeps improving slightly for a few more epochs.}

\begin{figure}[H]
     
     \includegraphics[width=0.66\textwidth]{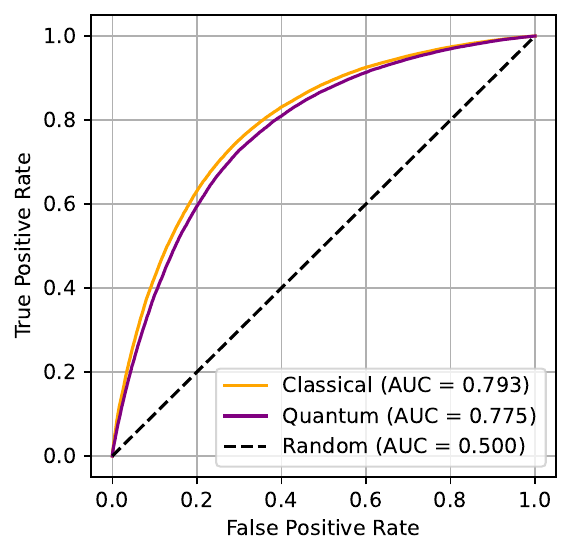}
     \caption{\added{Receiver Operating Characteristic (ROC) curves for the baseline classical ViT (orange line) and our hybrid QViT (purple line). The black dashed line represents the performance of a random~classifier.}}
     \label{fig:pr}
\end{figure}

With the parameters from the best epoch of each model, we compute the ROC curve and AUC score on the separate hold-out test set. We show the achieved test ROC curve and its AUC scores in Figure \ref{fig:pr}.
\vspace{-7pt}

\replaced{We observe that the proposed QViT results in almost the same ROC curve and obtains almost the same AUC score as the classical baseline,}{We observe that the proposed QViT obtains almost the same AUC as the classical baseline,} though it still lags by approximately two percentage points. We hypothesize that one potential reason for the slightly inferior performance of the quantum model is that it is harder for the optimizer to find good parameters within the numerically simulated VQCs. Alternatively, the proposed VQCs might lack the expressiveness required to match or exceed the performance of the classical model. Still, we note that the difference between both obtained metrics is quite~small.

\section{Conclusions}

In this work, we introduced a quantum-classical hybrid approach to vision transformers and applied it to the task of quark-gluon classification of sub-detector images from the CMS Open Data. \replaced{The novel element is the integration of variational quantum circuits within both the attention mechanism and the multi-layer perceptrons. The trained model was benchmarked against a classical vision transformer with the same hyperparameters and a similar number of trainable parameters and was found to have comparable performance.}{While the proposed model still performs slightly worse than its classical counterpart, the difference in the results is small.} The results achieved so far are encouraging and warrant future investigations. 

Moving forward, our plans include evaluating more hyperparameter configurations, assessing the impact of the number of training samples, and experimenting with data augmentation techniques \replaced{that}{which} have been shown to improve classical ViTs~\cite{bettervit,trainvit}, such as RandAugment~\cite{randaugment} and Mixup~\cite{mixup}. We also aim to explore different configurations for the VQCs, as well as evaluate the usage of data re-uploading~\cite{datareuploading} to check if we obtain a quantum advantage. Finally, we would also like to execute the VQCs on real quantum hardware to measure the performance of the proposed QViT in it, as well as to assess the robustness to quantum noise.

\added{Ideally, the progress in improving the performance of the ML and QML algorithms should be accompanied by progress in understanding the fundamental physics behind the hadronization of quarks and gluons. As a first step in this direction, one could use symbolic learning to obtain interpretable analytical formulas  that capture the decision-making of our trained classifiers~\cite{Dong:2022trn}.}

%All in all, the results achieved so far are encouraging and warrant future investigations. %Nevertheless, our research is still ongoing, therefore the findings are still preliminary.

%\paragraph{Reproducibility} The code and data we used to train and evaluate our models is available at \url{https://anonymous.4open.science/r/QuantumTransformers-anon/} (a link to a public GitHub repository will be provided on the non-anonymized version of the paper).

%%%%%%%%%%%%%%%%%%%%%%%%%%%%%%%%%%%%%%%%%%
\vspace{6pt} 

%%%%%%%%%%%%%%%%%%%%%%%%%%%%%%%%%%%%%%%%%%
%% optional
%\supplementary{The following supporting information can be downloaded at:  \linksupplementary{s1}, Figure S1: title; Table S1: title; Video S1: title.}


%%%%%%%%%%%%%%%%%%%%%%%%%%%%%%%%%%%%%%%%%%
\authorcontributions{
Conceptualization, M.C.C.; 
methodology, M.C.C., G.R.D., Z.D., R.T.F., S.G., D.J., K.K., T.M., K.T.M., K.M. and E.B.U.; 
software, M.C.C.; 
validation, M.C.C., G.R.D., Z.D., R.T.F., T.M. and E.B.U.; 
formal analysis, M.C.C.; 
investigation, M.C.C., G.R.D., Z.D., R.T.F., T.M. and E.B.U.; 
resources, M.C.C. and S.G.; 
data curation, G.R.D., S.G. and T.M.; 
writing---original draft preparation, M.C.C.; 
writing---review and editing, S.G., D.J., K.K., K.T.M. and K.M.; 
visualization, M.C.C.;
supervision, S.G., D.J., K.K., K.T.M. and K.M.; 
project administration, S.G., D.J., K.K., K.T.M. and K.M.; 
funding acquisition, S.G. 
All authors have read and agreed to the published version of the~manuscript.}

\funding{This research used resources of the National Energy Research Scientific Computing Center, a DOE Office of Science User Facility supported by the Office of Science of the U.S. Department of Energy under Contract No. DE-AC02-05CH11231 using NERSC award NERSC DDR-ERCAP0025759. SG is supported in part by the U.S. Department of Energy (DOE) under Award No. DE-SC0012447. KM is supported in part by the U.S. DOE award number DE-SC0022148. KK is supported in part by US DOE DE-SC0024407. CD is supported in part by the College of Liberal Arts and Sciences Research Fund at the University of Kansas. CD, RF, EU, MCC, and TM were participants in the 2023 Google Summer of Code.}

\institutionalreview{Not applicable.}

%\informedconsent{Any research article describing a study involving humans should contain this statement. Please add ``Informed consent was obtained from all subjects involved in the study.'' OR ``Patient consent was waived due to REASON (please provide a detailed justification).'' OR ``Not applicable'' for studies not involving humans. You might also choose to exclude this statement if the study did not involve humans.
%Written informed consent for publication must be obtained from participating patients who can be identified (including by the patients themselves). Please state ``Written informed consent has been obtained from the patient(s) to publish this paper'' if applicable.}

\dataavailability{
The code and data we used to train and evaluate our models are available at \url{https://github.com/ML4SCI/QMLHEP/tree/main/Quantum_Transformers_Mar\%C3\%A7al_Comajoan_Cara} \added{(accessed on 14 March 2024)}.} 

%\acknowledgments{In this section you can acknowledge any support given which is not covered by the author contribution or funding sections. This may include administrative and technical support, or donations in kind (e.g., materials used for experiments).}

\conflictsofinterest{The authors declare no conflicts of interest. The funders had no role in the design of the study; in the collection, analyses, or interpretation of data; in the writing of the manuscript; or in the decision to publish the results.} 

%%%%%%%%%%%%%%%%%%%%%%%%%%%%%%%%%%%%%%%%%%
%% Optional
%\sampleavailability{Samples of the compounds ... are available from the authors.}

%% Only for journal Encyclopedia
%\entrylink{The Link to this entry published on the encyclopedia platform.}
\newpage
\abbreviations{Abbreviations}{
The following abbreviations are used in this manuscript:\\

\noindent 
\begin{tabular}{@{}ll}
ALICE & A Large Ion Collider Experiment \\
ATLAS & A Toroidal LHC ApparatuS\\
AUC & Area Under the Curve \\
CERN & Conseil Europ\'{e}en pour la Recherche Nucl\'{e}aire\\
CMS & Compact Muon Solenoid (experiment)\\
CNOT & Controlled NOT\\
ECAL & electromagnetic calorimeter\\
GELU & Gaussian Error Linear Unit \\
HCAL & hadronic calorimeter\\
HL-LHC &High Luminosity Large Hadron Collider\\
LHC & Large Hadron Collider\\
LHCb & Large Hadron Collider beauty (experiment)\\
MDPI & Multidisciplinary Digital Publishing Institute\\
MHA & multi-head attention \\
MLP & multi-layer perceptron \\
QCD & Quantum Chromodynamics \\
QMHA & quantum multi-head attention \\
QML & quantum machine learning\\
QMLP & quantum multi-layer perceptron \\
QViT & quantum vision transformer\\
ReLU & Rectified Linear Unit \\
ROC & receiver operating characteristic\\
SGD & stochastic gradient descent \\
ViT & vision transformer\\
VQC & variational quantum circuit
\end{tabular}
}

%%%%%%%%%%%%%%%%%%%%%%%%%%%%%%%%%%%%%%%%%%
\begin{adjustwidth}{-\extralength}{0cm}

\reftitle{{References} %MDPI: Please DO NOT change/revert the form of references in Reference Section, they have been completed layout and ready for publication. Otherwise we cannot process to the next step. However, you may change the reference orders as necessary. Please NEVER use EndNote or other tools to rearrange the reference order. If it is hard to revise all the references into numerical order, we will help you to do this process. But please make sure every reference should be cited (not missing) and Reference cited in the main text match the reference in the Reference List. Please provide the detailed information if required in the comments below. Or please provide the website links and accessed date (Day Month Year) if you cannot provide detailed information.Very Important: References are not allowed to be added or deleted without reasons after the manuscript is accepted..
}

\PublishersNote{}
\end{adjustwidth}

\begin{thebibliography}{999}

\bibitem[CERN(2022)]{hllhc}
CERN.
\newblock {The HL-LHC Project}.  2022.
\newblock Available online:
\url{https://hilumilhc.web.cern.ch/content/hl-lhc-project} (accessed on 24 September
2023).

\bibitem[Amoroso et~al.(2021)]{HSFPhysicsEventGeneratorWG:2020gxw}
HSF Physics Event Generator WG; Valassi, A.; Yazgan, E.; McFayden, J.; Amoroso, S.; Bendavid, J.; Buckley, A.; Cacciari, M.; Childers, T.; Ciulli, V.; et al.
\newblock {Challenges in Monte Carlo Event Generator Software for
High-Luminosity LHC}.
\newblock {\em Comput. Softw. Big Sci.} {\bf 2021}, {\em 5},~{12}.
\newblock [\href{http://doi.org/10.1007/s41781-021-00055-1}{CrossRef}]

\bibitem[Arunachalam and de~Wolf(2017)]{qml1}
Arunachalam, S.; de~Wolf, R.
\newblock A Survey of Quantum Learning Theory. \emph{arXiv} \textbf{\boldmath{2017}}, arXiv:1701.06806. [\href{https://doi.org/10.48550/arXiv.1701.06806}{CrossRef}]
%\newblock {\url{https://doi.org/10.48550/arXiv.1701.06806}}.

\bibitem[Biamonte et~al.(2017)Biamonte, Wittek, Pancotti, Rebentrost, Wiebe,
and Lloyd]{qml2}
Biamonte, J.; Wittek, P.; Pancotti, N.; Rebentrost, P.; Wiebe, N.; Lloyd, S.
\newblock Quantum machine learning.
\newblock {\em Nature} {\bf 2017}, {\em 549},~195--202.
\newblock [\href{http://dx.doi.org/10.1038/nature23474}{CrossRef}] [\href{http://www.ncbi.nlm.nih.gov/pubmed/28905917}{PubMed}]

\bibitem[Schuld and Killoran(2019)]{hilbert}
Schuld, M.; Killoran, N.
\newblock Quantum Machine Learning in Feature Hilbert Spaces.
\newblock {\em Phys. Rev. Lett.} {\bf 2019}, {\em 122},~040504.
\newblock [\href{http://dx.doi.org/10.1103/PhysRevLett.122.040504}{CrossRef}] [\href{http://www.ncbi.nlm.nih.gov/pubmed/30768345}{PubMed}]

\bibitem[Mangini et~al.(2021)Mangini, Tacchino, Gerace, Bajoni, and
Macchiavello]{qml3}
Mangini, S.; Tacchino, F.; Gerace, D.; Bajoni, D.; Macchiavello, C.
\newblock Quantum computing models for artificial neural networks.
\newblock {\em Europhys. Lett.} {\bf 2021}, {\em 134},~10002.
\newblock [\href{http://dx.doi.org/10.1209/0295-5075/134/10002}{CrossRef}]

\bibitem[Liu et~al.(2021)Liu, Arunachalam, and Temme]{qml4}
Liu, Y.; Arunachalam, S.; Temme, K.
\newblock A rigorous and robust quantum speed-up in supervised machine
learning.
\newblock {\em Nat. Phys.} {\bf 2021}, {\em 17},~1013--1017.
\newblock [\href{http://dx.doi.org/10.1038/s41567-021-01287-z}{CrossRef}]

\bibitem[Huang et~al.(2022)Huang, Broughton, Cotler, Chen, Li, Mohseni, Neven,
Babbush, Kueng, Preskill, and McClean]{qml5}
Huang, H.Y.; Broughton, M.; Cotler, J.; Chen, S.; Li, J.; Mohseni, M.; Neven,
H.; Babbush, R.; Kueng, R.; Preskill, J.;  et~al.
\newblock Quantum advantage in learning from experiments.
\newblock {\em Science} {\bf 2022}, {\em 376},~1182--1186.
\newblock [\href{http://dx.doi.org/10.1126/science.abn7293}{CrossRef}]

\bibitem[Caro et~al.(2022)Caro, Huang, Cerezo, Sharma, Sornborger, Cincio, and
Coles]{qml6}
Caro, M.C.; Huang, H.Y.; Cerezo, M.; Sharma, K.; Sornborger, A.; Cincio, L.;
Coles, P.J.
\newblock Generalization in quantum machine learning from few training data.
\newblock {\em Nat. Commun.} {\bf 2022}, {\em 13},~4919.
\newblock [\href{http://dx.doi.org/10.1038/s41467-022-32550-3}{CrossRef}] [\href{http://www.ncbi.nlm.nih.gov/pubmed/35995777}{PubMed}]

\bibitem[Dong et~al.(2024)Dong, Comajoan~Cara, Dahale, Forestano, Gleyzer,
Justice, Kong, Magorsch, Matchev, Matcheva, and Unlu]{axioms13030188_Cosmos}
Dong, Z.; Comajoan~Cara, M.; Dahale, G.R.; Forestano, R.T.; Gleyzer, S.;
Justice, D.; Kong, K.; Magorsch, T.; Matchev, K.T.; Matcheva, K.;  et~al.
\newblock $Z_2\times Z_2$ Equivariant Quantum Neural Networks: Benchmarking
against Classical Neural Networks.
\newblock {\em Axioms} {\bf 2024}, {\em 13}, {188}
.
\newblock [\href{http://dx.doi.org/10.3390/axioms13030188}{CrossRef}]

\bibitem[Forestano et~al.(2024)Forestano, Comajoan~Cara, Dahale, Dong, Gleyzer,
Justice, Kong, Magorsch, Matchev, Matcheva, and Unlu]{axioms13030160_Roy}
Forestano, R.T.; Comajoan~Cara, M.; Dahale, G.R.; Dong, Z.; Gleyzer, S.;
Justice, D.; Kong, K.; Magorsch, T.; Matchev, K.T.; Matcheva, K.;  et~al.
\newblock A Comparison between Invariant and Equivariant Classical and Quantum
Graph Neural Networks.
\newblock {\em Axioms} {\bf 2024}, {\em 13}, {160}.
\newblock [\href{http://dx.doi.org/10.3390/axioms13030160}{CrossRef}]

\bibitem[Dosovitskiy et~al.(2021)Dosovitskiy, Beyer, Kolesnikov, Weissenborn,
Zhai, Unterthiner, Dehghani, Minderer, Heigold, Gelly, Uszkoreit, and
Houlsby]{vit}
Dosovitskiy, A.; Beyer, L.; Kolesnikov, A.; Weissenborn, D.; Zhai, X.;
Unterthiner, T.; Dehghani, M.; Minderer, M.; Heigold, G.; Gelly, S.;  et~al.
\newblock An Image is Worth 16 $\times$ 16 Words: Transformers for Image Recognition at
Scale.
\newblock In Proceedings of the International Conference on Learning
Representations,  {Online, 3--7 May} 2021.

\bibitem[Di~Sipio et~al.(2022)Di~Sipio, Huang, Chen, Mangini, and
Worring]{disipio}
Di~Sipio, R.; Huang, J.H.; Chen, S.Y.C.; Mangini, S.; Worring, M.
\newblock The Dawn of Quantum Natural Language Processing.
\newblock In Proceedings of the ICASSP 2022---2022 IEEE International
Conference on Acoustics, Speech and Signal Processing (ICASSP), {Singapore, 23--27 May} 2022; pp.
8612--8616.
\newblock [\href{http://dx.doi.org/10.1109/ICASSP43922.2022.9747675}{CrossRef}]

\bibitem[Vaswani et~al.(2017)Vaswani, Shazeer, Parmar, Uszkoreit, Jones, Gomez,
Kaiser, and Polosukhin]{transformer}
Vaswani, A.; Shazeer, N.; Parmar, N.; Uszkoreit, J.; Jones, L.; Gomez, A.N.;
Kaiser, L.u.; Polosukhin, I.
\newblock Attention is All you Need.
\newblock In Proceedings of the Advances in Neural Information Processing
Systems, {Long Beach, CA, USA, 4--9 December 2017}; Guyon, I., Luxburg, U.V., Bengio, S., Wallach, H., Fergus, R.,
Vishwanathan, S., Garnett, R., Eds.; Curran Associates, Inc.: {New York, NY, USA,} 2017; Volume~30.

\bibitem[Li et~al.(2022)Li, Zhao, and Wang]{li}
Li, G.; Zhao, X.; Wang, X.
\newblock Quantum Self-Attention Neural Networks for Text Classification. \emph{arXiv} \textbf{\boldmath{2022}}, arXiv:2205.05625.
\newblock [\href{https://doi.org/10.48550/arXiv.2205.05625}{CrossRef}]

\bibitem[Cherrat et~al.(2023)Cherrat, Kerenidis, Mathur, Landman, Strahm, and
Li]{cherrat}
Cherrat, E.A.; Kerenidis, I.; Mathur, N.; Landman, J.; Strahm, M.C.; Li, Y.Y.
\newblock Quantum Vision Transformers.
\newblock {\em Quantum} {\bf 2024}, {\em 8}, 1265.
\newblock [\href{http://dx.doi.org/10.22331/q-2024-02-22-1265}{CrossRef}]
%MDPI: We are sorry that we didn't find any references that match the information in this entry. Please provide more information indicate what type of article is it or any accessible method to it, such as book (please provide the name and location of the publisher); online resource (please provide the URL of the website and the date it was accessed (Date Month Year)); or journal article (please provide the name of the journal, the year and volume in which it was published, and the page number). Please refer to https://www.mdpi.com/authors/references for full reference formatting guides.


\bibitem[Unlu et~al.(2024)Unlu, Comajoan~Cara, Dahale, Dong, Forestano,
Gleyzer, Justice, Kong, Magorsch, Matchev, and Matcheva]{axioms13030187_Eyup}
Unlu, E.B.; Comajoan~Cara, M.; Dahale, G.R.; Dong, Z.; Forestano, R.T.;
Gleyzer, S.; Justice, D.; Kong, K.; Magorsch, T.; Matchev, K.T.;  et~al.
\newblock Hybrid Quantum Vision Transformers for Event Classification in High
Energy Physics.
\newblock {\em Axioms} {\bf 2024}, {\em 13}, {187}.
\newblock [\href{http://dx.doi.org/10.3390/axioms13030187}{CrossRef}]

\bibitem[Kollias et~al.(2023)Kollias, Kalantzis, Salonidis, and Ubaru]{qgt}
Kollias, G.; Kalantzis, V.; Salonidis, T.; Ubaru, S.
\newblock Quantum Graph Transformers.
\newblock In Proceedings of the ICASSP 2023---2023 IEEE International
Conference on Acoustics, Speech and Signal Processing (ICASSP), {Rhodes Island, Greece, 4--10 June} 2023; pp.
1--5.
\newblock [\href{http://dx.doi.org/10.1109/ICASSP49357.2023.10096345}{CrossRef}]

\bibitem[CERN(2023)]{cmsopendata}
CERN.
\newblock {CMS Open Data}.  2023.
\newblock Available online: \url{http://opendata.cern.ch/docs/about-cms}
(accessed on 24 September 2023).

\bibitem[{The ATLAS Collaboration}(2017)]{qgatlas}
{The ATLAS Collaboration}.
\newblock \emph{Quark versus Gluon Jet Tagging Using Jet Images with the ATLAS
Detector};
\newblock Technical Report; CERN:~Geneva, {Switzerland,} 2017.
\newblock Available online: \url{https://cds.cern.ch/record/2275641}
(accessed on 12 May 2024).

\bibitem[{The CMS Collaboration}(2017)]{qgcms}
{The CMS Collaboration}.
\newblock New Developments for Jet Substructure Reconstruction in CMS. 2017.  Available online: \url{https://cds.cern.ch/record/2275226}  ({accessed on 8 May 2024}
).

\bibitem[Cheng(2018)]{qg1}
Cheng, T.
\newblock Recursive Neural Networks in Quark/Gluon Tagging.
\newblock {\em Comput. Softw. Big Sci.} {\bf 2018}, {\em 2}, {3}.
\newblock [\href{http://dx.doi.org/10.1007/s41781-018-0007-y}{CrossRef}]

\bibitem[Louppe et~al.(2019)Louppe, Cho, Becot, and Cranmer]{qg2}
Louppe, G.; Cho, K.; Becot, C.; Cranmer, K.
\newblock QCD-aware recursive neural networks for jet physics.
\newblock {\em J. High Energy Phys.} {\bf 2019}, {\em 2019},~57.
\newblock [\href{http://dx.doi.org/10.1007/JHEP01(2019)057}{CrossRef}]

\bibitem[Andrews et~al.(2020)Andrews, Alison, An, Burkle, Gleyzer, Narain,
Paulini, Poczos, and Usai]{qggleyzer}
Andrews, M.; Alison, J.; An, S.; Burkle, B.; Gleyzer, S.; Narain, M.; Paulini,
M.; Poczos, B.; Usai, E.
\newblock End-to-end jet classification of quarks and gluons with the {CMS}
Open Data.
\newblock {\em Nucl. Instrum. Methods Phys. Res. Sect. A Accel. Spectrometers Detect. Assoc. Equip.} {\bf 2020},
{\em 977},~164304.
\newblock [\href{http://dx.doi.org/10.1016/j.nima.2020.164304}{CrossRef}]

\bibitem[LeCun et~al.(2015)LeCun, Bengio, and Hinton]{lbh}
LeCun, Y.; Bengio, Y.; Hinton, G.
\newblock Deep learning.
\newblock {\em Nature} {\bf 2015}, {\em 521},~436--444.
\newblock [\href{http://dx.doi.org/10.1038/nature14539}{CrossRef}]

\bibitem[Bishop and Bishop(2023)]{bishop}
Bishop, C.M.; Bishop, H.
\newblock {\em Deep Learning}, 1st ed.; Springer: Cham, Switzerland,  2023.

\bibitem[Schmidhuber(2022)]{sh}
Schmidhuber, J.
\newblock Annotated History of Modern AI and Deep Learning. \emph{arXiv} \textbf{\boldmath{2022}}, arXiv:2212.11279.

\bibitem[Fukushima(1969)]{relu1}
Fukushima, K.
\newblock Visual Feature Extraction by a Multilayered Network of Analog
Threshold Elements.
\newblock {\em IEEE Trans. Syst. Sci. Cybern.} {\bf
1969}, {\em 5},~322--333.
\newblock [\href{http://dx.doi.org/10.1109/TSSC.1969.300225}{CrossRef}]

\bibitem[Glorot et~al.(2011)Glorot, Bordes, and Bengio]{relu2}
Glorot, X.; Bordes, A.; Bengio, Y.
\newblock Deep sparse rectifier neural networks.
\newblock In Proceedings of the Fourteenth International
Conference on Artificial Intelligence and Statistics. JMLR Workshop and
Conference Proceedings, {Fort Lauderdale, FL, USA, 11--13 April}  2011; pp. 315--323.

\bibitem[Hendrycks and Gimpel(2016)]{gelu}
Hendrycks, D.; Gimpel, K.
\newblock Gaussian Error Linear Units (GELUs). \emph{arXiv} \textbf{\boldmath{2016}}, arXiv:1606.08415.

\bibitem[Ba et~al.(2016)Ba, Kiros, and Hinton]{layernorm}
Ba, J.L.; Kiros, J.R.; Hinton, G.E.
\newblock Layer normalization.
\newblock {\em arXiv}  {\bf 2016}, arXiv:1607.06450.

\bibitem[He et~al.(2015)He, Zhang, Ren, and Sun]{resnet}
He, K.; Zhang, X.; Ren, S.; Sun, J.
\newblock Deep Residual Learning for Image Recognition. \emph{arXiv}  \textbf{\boldmath{2015}}, arXiv:1512.03385.

\bibitem[Beyer et~al.(2022)Beyer, Zhai, and Kolesnikov]{bettervit}
Beyer, L.; Zhai, X.; Kolesnikov, A.
\newblock Better plain ViT baselines for ImageNet-1k. \emph{arXiv} {\bf 2022}, arXiv:2205.01580.
\newblock [\href{https://doi.org/10.48550/arXiv.2205.01580}{CrossRef}]

\bibitem[Rebentrost et~al.(2014)Rebentrost, Mohseni, and Lloyd]{qsvm}
Rebentrost, P.; Mohseni, M.; Lloyd, S.
\newblock Quantum Support Vector Machine for Big Data Classification.
\newblock {\em Phys. Rev. Lett.} {\bf 2014}, {\em 113},~{130503}.
\newblock [\href{http://dx.doi.org/10.1103/physrevlett.113.130503}{CrossRef}]

\bibitem[Wiebe et~al.(2015)Wiebe, Kapoor, and Svore]{qknn}
Wiebe, N.; Kapoor, A.; Svore, K.M.
\newblock Quantum Algorithms for Nearest-Neighbor Methods for Supervised and
Unsupervised Learning.
\newblock {\em Quantum Inf. Comput.} {\bf 2015}, {\em 15},~316–356.
%\newblock [\href{http://dx.doi.org/10.5555/2871393.2871400}{CrossRef}]

\bibitem[Johri et~al.(2021)Johri, Debnath, Mocherla, Singk, Prakash, Kim, and
Kerenidis]{qnc}
Johri, S.; Debnath, S.; Mocherla, A.; Singk, A.; Prakash, A.; Kim, J.;
Kerenidis, I.
\newblock Nearest centroid classification on a trapped ion quantum computer.
\newblock {\em npj Quantum Inf.} {\bf 2021}, {\em 7},~122.
\newblock [\href{http://dx.doi.org/10.1038/s41534-021-00456-5}{CrossRef}]

\bibitem[Franceschini et~al.(2023)Franceschini, Kim, Kong, Matchev, Park, and
Shyamsundar]{Franceschini:2022vck}
Franceschini, R.; Kim, D.; Kong, K.; Matchev, K.T.; Park, M.; Shyamsundar, P.
\newblock {Kinematic variables and feature engineering for particle
phenomenology}.
\newblock {\em Rev. Mod. Phys.} {\bf 2023}, {\em 95},~045004.
\newblock [\href{http://dx.doi.org/10.1103/RevModPhys.95.045004}{CrossRef}]

\bibitem[Ellis et~al.(2011)Ellis, Stirling, and Webber]{Ellis:1996mzs}
Ellis, R.K.; Stirling, W.J.; Webber, B.R.
\newblock {\em {QCD and Collider Physics}}; Cambridge University Press: {Cambdrige, UK},
2011; Volume~8.
\newblock [\href{http://dx.doi.org/10.1017/CBO9780511628788}{CrossRef}]

\bibitem[Salam(2010)]{Salam:2010nqg}
Salam, G.P.
\newblock {Towards Jetography}.
\newblock {\em Eur. Phys. J. C} {\bf 2010}, {\em 67},~637--686.
\newblock [\href{http://dx.doi.org/10.1140/epjc/s10052-010-1314-6}{CrossRef}]

\bibitem[Larkoski et~al.(2020)Larkoski, Moult, and Nachman]{Larkoski:2017jix}
Larkoski, A.J.; Moult, I.; Nachman, B.
\newblock {Jet Substructure at the Large Hadron Collider: A Review of Recent
Advances in Theory and Machine Learning}.
\newblock {\em Phys. Rept.} {\bf 2020}, {\em 841},~1--63.
\newblock [\href{http://dx.doi.org/10.1016/j.physrep.2019.11.001}{CrossRef}]

\bibitem[Kogler et~al.(2019)]{Kogler:2018hem}
Kogler, R.; {Nachman, B.; Schmidt, A.; Asquith, L.; Winkels, E.; Campanelli, M.; Delitzsch, C.; Harris, P.; Hinzmann, A.; Kar, D.;}  {et~al.}
\newblock {Jet Substructure at the Large Hadron Collider: Experimental Review}.
\newblock {\em Rev. Mod. Phys.} {\bf 2019}, {\em 91},~045003.
\newblock [\href{http://dx.doi.org/10.1103/RevModPhys.91.045003}{CrossRef}]

\bibitem[Marzani et~al.(2019)Marzani, Soyez, and Spannowsky]{Marzani:2019hun}
Marzani, S.; Soyez, G.; Spannowsky, M.
\newblock {\em {Looking Inside Jets: An Introduction to Jet Substructure and
Boosted-Object Phenomenology}}; Springer: Berlin/Heidelberg, Germany,  2019; Volume 958.
\newblock [\href{http://dx.doi.org/10.1007/978-3-030-15709-8}{CrossRef}]

\bibitem[Feickert and Nachman(2021)]{Feickert:2021ajf}
Feickert, M.; Nachman, B.
\newblock {A Living Review of Machine Learning for Particle Physics}. \emph{arXiv} {\bf
2021}, arXiv:2102.02770.

\bibitem[Guest et~al.(2018)Guest, Cranmer, and Whiteson]{Guest:2018yhq}
Guest, D.; Cranmer, K.; Whiteson, D.
\newblock {Deep Learning and its Application to LHC Physics}.
\newblock {\em Ann. Rev. Nucl. Part. Sci.} {\bf 2018}, {\em 68},~161--181.
\newblock [\href{http://dx.doi.org/10.1146/annurev-nucl-101917-021019}{CrossRef}]

\bibitem[Albertsson et~al.(2018)]{Albertsson:2018maf}
Albertsson, K.; Altoe, P.; Anderson, D.; Andrews, M.; Araque Espinosa, J.P.; Aurisano, A.; Basara, L.; Bevan, A.; Bhimji, W.; Bonacorsi, D.; et al.
\newblock {Machine Learning in High Energy Physics Community White Paper}.
\newblock {\em J. Phys. Conf. Ser.} {\bf 2018}, {\em 1085},~022008.
\newblock [\href{http://dx.doi.org/10.1088/1742-6596/1085/2/022008}{CrossRef}]

\bibitem[Radovic et~al.(2018)Radovic, Williams, Rousseau, Kagan, Bonacorsi,
Himmel, Aurisano, Terao, and Wongjirad]{Radovic:2018dip}
Radovic, A.; Williams, M.; Rousseau, D.; Kagan, M.; Bonacorsi, D.; Himmel, A.;
Aurisano, A.; Terao, K.; Wongjirad, T.
\newblock {Machine learning at the energy and intensity frontiers of particle
physics}.
\newblock {\em Nature} {\bf 2018}, {\em 560},~41--48.
\newblock [\href{http://dx.doi.org/10.1038/s41586-018-0361-2}{CrossRef}] [\href{http://www.ncbi.nlm.nih.gov/pubmed/30068955}{PubMed}]

\bibitem[Carleo et~al.(2019)Carleo, Cirac, Cranmer, Daudet, Schuld, Tishby,
Vogt-Maranto, and Zdeborov\'a]{Carleo:2019ptp}
Carleo, G.; Cirac, I.; Cranmer, K.; Daudet, L.; Schuld, M.; Tishby, N.;
Vogt-Maranto, L.; Zdeborov\'a, L.
\newblock {Machine learning and the physical sciences}.
\newblock {\em Rev. Mod. Phys.} {\bf 2019}, {\em 91},~045002.
\newblock [\href{http://dx.doi.org/10.1103/RevModPhys.91.045002}{CrossRef}]

\bibitem[Bourilkov(2020)]{Bourilkov:2019yoi}
Bourilkov, D.
\newblock {Machine and Deep Learning Applications in Particle Physics}.
\newblock {\em Int. J. Mod. Phys. A} {\bf 2020}, {\em 34},~1930019.
\newblock [\href{http://dx.doi.org/10.1142/S0217751X19300199}{CrossRef}]

\bibitem[Schwartz(2021)]{Schwartz:2021ftp}
Schwartz, M.D.
\newblock {Modern Machine Learning and Particle Physics.} \emph{arXiv} {\bf 2021}, arXiv:2103.12226.
\newblock [\href{https://doi.org/10.1162/99608f92.beeb1183}{CrossRef}]

\bibitem[Karagiorgi et~al.(2021)Karagiorgi, Kasieczka, Kravitz, Nachman, and
Shih]{Karagiorgi:2021ngt}
Karagiorgi, G.; Kasieczka, G.; Kravitz, S.; Nachman, B.; Shih, D.
\newblock {Machine Learning in the Search for New Fundamental Physics.} \emph{arXiv} {\bf
2021}, arXiv:2112.03769.

\bibitem[Boehnlein et~al.(2022)]{Boehnlein:2021eym}
Boehnlein, A.; Diefenthaler, M.; Sato, N.; Schram, M.; Ziegler, V.; Fanelli, C.; Hjorth-Jensen, M.; Horn, T.; Kuchera, M.P.; Lee, D.; et~al.
\newblock {Colloquium: Machine learning in nuclear physics}.
\newblock {\em Rev. Mod. Phys.} {\bf 2022}, {\em 94},~031003.
\newblock [\href{http://dx.doi.org/10.1103/RevModPhys.94.031003}{CrossRef}]

\bibitem[Shanahan et~al.(2022)]{Shanahan:2022ifi}
Shanahan, P.; Terao, K.; Whiteson, D.
\newblock {Snowmass 2021 Computational Frontier CompF03 Topical Group Report:
Machine Learning.} \emph{arXiv} {\bf 2022}, arXiv:2209.07559.

\bibitem[Chatrchyan et~al.(2008)]{CMS:2008xjf}
Collaboration, C.M.; Chatrchyan, S.; Hmayakyan, G.; Khachatryan, V.; Sirunyan, A.M.; Adam, W.; Bauer, T.; Bergauer, T.; Bergauer, H.; Dragicevic, M.; et al.
\newblock {The CMS Experiment at the CERN LHC}.
\newblock {\em JINST} {\bf 2008}, {\em 3},~S08004.
\newblock [\href{http://dx.doi.org/10.1088/1748-0221/3/08/S08004}{CrossRef}]

\bibitem[Chatrchyan et~al.(2014)]{CMS:2014pgm}
CMS Collaboration.
\newblock {Description and performance of track and primary-vertex
reconstruction with the CMS tracker}.
\newblock {\em JINST} {\bf 2014}, {\em 9},~P10009.
\newblock [\href{http://dx.doi.org/10.1088/1748-0221/9/10/P10009}{CrossRef}]

\bibitem[Chatrchyan et~al.(2013)]{CMS:2013lxn}
CMS Collaboration.
\newblock {Energy Calibration and Resolution of the CMS Electromagnetic
Calorimeter in $pp$ Collisions at $\sqrt{s} = 7$ TeV}.
\newblock {\em JINST} {\bf 2013}, {\em 8},~P09009.
\newblock [\href{http://dx.doi.org/10.1088/1748-0221/8/09/P09009}{CrossRef}]

\bibitem[Abdullin et~al.(2008{\natexlab{a}})]{CMSHCAL:2007zcq}
Abdullin, S.; Abramov, V.; Acharya, B.; Adams, M.; Akchurin, N.; Akgun, U.; Anderson, E.W.; Antchev, G.; Ayan, S.; Aydin, S.; et al.
\newblock {Design, performance, and calibration of CMS hadron-barrel
calorimeter wedges}.
\newblock {\em Eur. Phys. J. C} {\bf 2008}, {\em 55},~159--171.
\newblock [\href{http://dx.doi.org/10.1140/epjc/s10052-008-0573-y}{CrossRef}]

\bibitem[Abdullin et~al.(2008{\natexlab{b}})]{CMSHCAL:2008fum}
Abdullin, S.; Abramov, V.; Acharya, B.; Adams, M.; Akchurin, N.; Akgun, U.; Anderson, E.W.; Antchev, G.; Arcidy, M.
\newblock {Design, performance, and calibration of the CMS Hadron-outer
calorimeter}.
\newblock {\em Eur. Phys. J. C} {\bf 2008}, {\em 57},~653--663.
\newblock [\href{http://dx.doi.org/10.1140/epjc/s10052-008-0756-6}{CrossRef}]

\bibitem[CMS()]{CMS_Coordinate_System}
CMS Coordinate System.
\newblock  Available online: \url{https://tikz.net/axis3d_cms/} (accessed on 6 March 2024).

\bibitem[Herrmann et~al.(2023)Herrmann, Arya, Doherty, Mingare, Pillay, Preis,
and Prestel]{Herrmann_2023}
Herrmann, N.; Arya, D.; Doherty, M.W.; Mingare, A.; Pillay, J.C.; Preis, F.;
Prestel, S.
\newblock Quantum utility---Definition and assessment of a practical quantum
advantage.
\newblock In Proceedings of the 2023 IEEE International Conference on Quantum
Software, {Chicago, IL, USA, 2--8 July} 
2023; pp. 162--174.
\newblock [\href{http://dx.doi.org/10.1109/qsw59989.2023.00028}{CrossRef}]

\bibitem[Loshchilov and Hutter(2019)]{adamw}
Loshchilov, I.; Hutter, F.
\newblock Decoupled Weight Decay Regularization. \emph{arXiv} {\bf 2019}, arXiv:1711.05101.
\newblock [\href{https://doi.org/10.48550/arXiv.1711.05101}{CrossRef}]

\bibitem[Loshchilov and Hutter(2017)]{cosine}
Loshchilov, I.; Hutter, F.
\newblock SGDR: Stochastic Gradient Descent with Warm Restarts. \emph{arXiv} {\bf 2017}, arXiv:1608.03983.

\bibitem[Bradbury et~al.(2023)Bradbury, Frostig, Hawkins, Johnson, Leary,
Maclaurin, Necula, Paszke, Vander{P}las, Wanderman-{M}ilne, and Zhang]{jax}
Bradbury, J.; Frostig, R.; Hawkins, P.; Johnson, M.J.; Leary, C.; Maclaurin,
D.; Necula, G.; Paszke, A.; Vander{P}las, J.; Wanderman-{M}ilne, S.;  et~al.
\newblock {JAX}: Composable Transformations of {P}ython+{N}um{P}y Programs.
2023.
\newblock Available online: \url{http://github.com/google/jax} (accessed on 24 September
2023).

\bibitem[Heek et~al.(2023)Heek, Levskaya, Oliver, Ritter, Rondepierre, Steiner,
and van {Z}ee]{flax}
Heek, J.; Levskaya, A.; Oliver, A.; Ritter, M.; Rondepierre, B.; Steiner, A.;
van {Z}ee, M.
\newblock {F}lax: A Neural Network Library and Ecosystem for {JAX}.  2023.
\newblock Available online: \url{http://github.com/google/flax} (accessed on 24
September 2023).

\bibitem[Zhang et~al.(2023)Zhang, Allcock, Wan, Liu, Sun, Yu, Yang, Qiu, Ye,
Chen, Lee, Zheng, Jian, Yao, Hsieh, and Zhang]{tensorcircuit}
Zhang, S.X.; Allcock, J.; Wan, Z.Q.; Liu, S.; Sun, J.; Yu, H.; Yang, X.H.; Qiu,
J.; Ye, Z.; Chen, Y.Q.;  et~al.
\newblock Tensor{C}ircuit: A {Q}uantum {S}oftware {F}ramework for the {NISQ}
{E}ra.
\newblock {\em {Quantum}} {\bf 2023}, {\em 7},~912.
\newblock [\href{http://dx.doi.org/10.22331/q-2023-02-02-912}{CrossRef}]

\bibitem[Steiner et~al.(2022)Steiner, Kolesnikov, Zhai, Wightman, Uszkoreit,
and Beyer]{trainvit}
Steiner, A.; Kolesnikov, A.; Zhai, X.; Wightman, R.; Uszkoreit, J.; Beyer, L.
\newblock How to train your ViT? Data, Augmentation, and Regularization in
Vision Transformers. \emph{arXiv} {\bf 2022}, arXiv:2106.10270.
\newblock [\href{https://doi.org/10.48550/arXiv.2106.10270}{CrossRef}]

\bibitem[Cubuk et~al.(2020)Cubuk, Zoph, Shlens, and Le]{randaugment}
Cubuk, E.D.; Zoph, B.; Shlens, J.; Le, Q.
\newblock RandAugment: Practical Automated Data Augmentation with a Reduced
Search Space.
\newblock In Proceedings of the Advances in Neural Information Processing
Systems, {Online, 6--12 December 2020}; Larochelle, H., Ranzato, M., Hadsell, R., Balcan, M., Lin, H., Eds.;
Curran Associates, Inc.: {Red Hook, NY, USA,}
2020; Volume~33, pp. 18613--18624.

\bibitem[Zhang et~al.(2018)Zhang, Cisse, Dauphin, and Lopez-Paz]{mixup}
Zhang, H.; Cisse, M.; Dauphin, Y.N.; Lopez-Paz, D.
\newblock mixup: Beyond Empirical Risk Minimization.
\newblock In Proceedings of the International Conference on Learning
Representations,  {Vancouver, BC, Canada, 30 April--3 May} 2018.
%\newpage
\bibitem[P{\'{e}}rez-Salinas et~al.(2020)P{\'{e}}rez-Salinas, Cervera-Lierta,
Gil-Fuster, and Latorre]{datareuploading}
P{\'{e}}rez-Salinas, A.; Cervera-Lierta, A.; Gil-Fuster, E.; Latorre, J.I.
\newblock Data re-uploading for a universal quantum classifier.
\newblock {\em {Quantum}} {\bf 2020}, {\em 4},~226.
\newblock [\href{http://dx.doi.org/10.22331/q-2020-02-06-226}{CrossRef}]

\bibitem[Dong et~al.(2023)Dong, Kong, Matchev, and Matcheva]{Dong:2022trn}
Dong, Z.; Kong, K.; Matchev, K.T.; Matcheva, K.
\newblock {Is the machine smarter than the theorist: Deriving formulas for
particle kinematics with symbolic regression}.
\newblock {\em Phys. Rev. D} {\bf 2023}, {\em 107},~055018.
\newblock [\href{http://dx.doi.org/10.1103/PhysRevD.107.055018}{CrossRef}]

\end{thebibliography}
\end{document}